# New Stable and Fast Ring-Polymer Molecular Dynamics for Calculating Bimolecular Rate Coefficients with Example of OH + CH$_4$

Xiongfei Gui[a], Wenbin Fan[b], Jiace Sun[c], and Yongle Li[*a]

a Department of Physics, International Center of Quantum and Molecular Structures, and Shanghai Frontiers Science Center of Quantum and Superconducting Matter States, Shanghai 200444, China

b Department of Chemistry, Fudan University, Shanghai 200433, China

c Division of Chemistry and Chemical Engineering, California Institute of Technology, Pasadena, California 91125, United States

* yongleli@shu.edu.cn

**ABSTRACT:** The accurate and efficient calculation of the rate coefficients of chemical reactions is a key issue in the research of chemical dynamics. In this work, by applying the dimension-free ultra-stable Cayley propagator, the thermal rate coefficients of a prototypic high dimensional chemical reaction OH + CH$_4$ → H$_2$O + CH$_3$ in the temperature range of 200 K to 1500 K are investigated with ring polymer molecular dynamics (RPMD), on a highly accurate full-dimensional potential energy surface. Kinetic isotope effects (KIEs) for three isotopologues of the title reaction are also studied. The results demonstrate excellent agreement with experimental data, even in the deep tunneling region. Especially, the Cayley propagator shows high calculation efficiency with little loss of accuracy. The present results confirmed the applicability of the RPMD method, particularly the speed-up by Cayley propagator, in theoretical calculations of bimolecular reaction rates.

## 1. Introduction

The rate coefficients of radical–molecule reactions attract wide interest in both atmospheric and combustion chemistry, because of their importance in modeling the kinetics and revealing the mechanism. In particular, the kinetic isotope effects (KIEs) calculated for the same reaction with different isotopes shed light on the reaction mechanism and quantum effects. However, the precise measurement of rate coefficients and KIEs is often a difficult task and the data are not always available.[1] So, theoretical calculations are necessary, since they not only offer reliable predictions of rate coefficients and KIEs but also provide microscopic insight into the reaction mechanism.

Currently, there are several categories of methodologies in calculating rate coefficients. The first approach is accurate quantum dynamics (QD) with wave packets, but it suffers from high computational costs. The second method is quasi-classical trajectory (QCT), which is fast, but incapable of treating quantum effects such as tunneling. The third method is the transition state theory (TST), which requires information only near the barrier. However, TST treats the tunneling approximately and may miss important recrossing dynamics near the barrier. The last one is ring-polymer molecular dynamics (RPMD), which achieves a good balance between accuracy and efficiency. In the past decade, RPMD has evidenced a myriad of successful applications for calculating bimolecular reaction rate coefficients.[2, 3]

The reaction of $OH + CH_4 \rightarrow H_2O + CH_3$ is a prototypical polyatomic reaction with 15 degrees of freedom (DOF). It is a major channel for methane removal in the atmosphere at low temperatures and is also a key step in combustion chemistry at high temperatures. Extensive experiments have been performed on title reaction[4-18], yielding thermal rate coefficients, KIEs, and even state-resolved rates over a large range of temperature. However, full dimensional QD calculations have not been reported yet due to the high dimensionality. On the other hand, due to the multidimensional tunneling of the transferring H atom, TST and QCT calculations could contain substantial inaccuracies at low temperatures.

Recently, a high accuracy full-dimensional neural-network potential energy surface (PIP-NN PES)[19] has been reported. With variational transition state theory (VTST) and quasi-classical trajectory (QCT), Li and Guo calculated the rate coefficients and KIEs and compare them with experimental data at temperatures ranging from 200 K to 1000 K[20]. Relatively large errors were discovered when the temperature drops below 300 K. This discrepancy presumably stems from the lack of multi-dimensional tunneling in both VTST and QCT. More specifically, the anharmonicity leads to a lower rate, and the lack of recrossing leads to a higher result, a combination of both caused error cancellation at high temperatures. This cancellation makes the calculated rate coefficients close to the experiments but the KIEs contain large errors. When temperature falls into the deep tunneling region, tunneling, recrossing, anharmonicity and zero-point energy (ZPE) effects become substantial, resulting in larger deviations from experiments.

Considering the above problems, a proper method based on a correct description of rate coefficients should reproduce both quantum and recrossing effects. The RPMD based on the isomorphism between a set of ring-polymers in extended phase space and the original quantum system, using real-time dynamics to calculate time correlation functions for obtaining the rate coefficients[21, 22], is a perfect candidate.

Previously, RPMD has been used for the investigation of title reaction on a less accurate analytical PES,[3, 23] where the rate coefficients at low temperatures were severely overestimated by one order of magnitude. Given the reliability of RPMD in other bimolecular reactions, such deficiency is presumably attributable to the deficiency of the PES. In this work, we report RPMD calculations using the accurate PIP-NN PES. However, the PIP-NN PES is much more computationally expensive due to its complex functional form, hence numerical acceleration of the RPMD propagation becomes critical.[24, 25] Recently, the Cayley propagator[23, 24] was proposed for RPMD calculation to endow strong stability, ergodicity, and dimensionality freedom. Such properties imply the ability to speed up the calculation by employing a larger time step without loss of accuracy. In this work, we deploy Cayley propagator in our RPMD calculations.

We organized this work as follows. The method is summarized in Sec. 2. All results are presented and discussed in Sec. 3, and the conclusion is drawn in Sec. 4.

## 2 Methods

### 2.1 RPMD

The theoretical method is summarized thoroughly elsewhere[2, 26], so we only give a brief description. For the title reaction, the quantum Hamiltonian is:

$$\hat{H} = \sum_{i=1}^{7} \frac{|\hat{p}_i|^2}{2m_i} + V(\hat{q}_1, \cdots, \hat{q}_7), \quad (1)$$

where $\hat{p}_i$ and $\hat{q}_i$ are the momentum and position operators of the $i$ th atom with its mass $m_i$. The RPMD Hamiltonian can be written down based on the isomorphism between the quantum system and the corresponding classical ring-polymer system:

$$\hat{H}_P(p,q) = \sum_{j=1}^{P} \left\{ \sum_{i=1}^{7} \left[ \frac{|p_i^{(j)}|^2}{2m_i} + \frac{1}{2} m_i \omega_n^2 \left( q_i^{(j)} - q_i^{(j-1)} \right)^2 \right] + V\left( q_1^{(j)}, \cdots, q_7^{(j)} \right) \right\}, (2)$$

here each atom is presented as a ring formed by a set of $P$ beads. The cyclic boundary is used: $q_0 = q_P$, and the force constant of harmonic potential connecting adjacent beads is $\omega_P = (\beta_P \hbar)^{-1}$, with the reciprocal temperature of the system $\beta_P = \beta / P$ and $\beta = 1/(k_B T)$.

Making use of the Bennett–Chandler factorization[27, 28], the RPMD rate coefficient is given as[29]

$$k_{\text{RPMD}}(T) = k_{\text{QTST}}(T;\xi^{\ddagger})\kappa(t \to \infty;\xi^{\ddagger})f(T). \qquad (3)$$

The first term $k_{\text{QTST}}(T;\xi^{\ddagger})$ is the quantum transition state theory (QTST) rate coefficients, which is evaluated at the peak of the free energy barrier in the reaction coordinate $\xi^{\ddagger}$.

The reaction coordinate (RC) of the title reaction is constructed from a pair of dividing surfaces defined in terms of the centroid of the ring polymer. The first dividing surface is defined by a distance $R_{\infty}$, which is the distance between the centers of mass (COM) of both reactants OH and CH$_4$, and is large enough making the interaction between them negligible:[30, 31]

$$s_0(\mathbf{q}) = R_{\infty} - \left| \bar{\mathbf{r}}_{\text{OH}}^{(\text{COM})} - \bar{\mathbf{r}}_{\text{CH}_4}^{(\text{COM})} \right| \qquad (4)$$

where $\bar{\mathbf{r}}_{\text{AB}}$ is the centroid position vector of group AB's COM $\bar{\mathbf{r}}_{\text{AB}} = \bar{\mathbf{r}}_{\text{B}} - \bar{\mathbf{r}}_{\text{A}}$, and $\bar{\mathbf{r}}_{\text{A}} = \dfrac{\sum_{i \in A} m_i \bar{\mathbf{r}}_i}{\sum_{i \in A} m_i}$.

And the centroid position vector is obtained from averaging the position vectors of beads $\bar{\mathbf{r}}_i = \dfrac{1}{P}\sum_{j=1}^{P} \mathbf{r}_i^{(j)}$. The second dividing surface is located at the transition state region,[30, 31]

$$s_1(\mathbf{q}) = \max_i \left[ \left( \left|\bar{\mathbf{r}}_{\text{CH}_i}\right| - r_{\text{CH}_i}^{\ddagger} \right) - \left( \left|\bar{\mathbf{r}}_{\text{OH}_i}\right| - r_{\text{OH}_i}^{\ddagger} \right) \right], \qquad (5)$$

where H$_i$ denotes any of the hydrogen atoms belonging to methane and $r_{\text{AB}}^{\ddagger}$ is the distance between atom A and B at the transition state. With both of the dividing surfaces, one can construct the dimensionless reaction coordinate used in RPMD rate coefficient calculation.

$$\xi(\bar{\mathbf{q}}) = \dfrac{s_0(\bar{\mathbf{q}})}{s_0(\bar{\mathbf{q}}) - s_1(\bar{\mathbf{q}})} \qquad (6)$$

In practice, $k_{\text{QTST}}(T;\xi^{\ddagger})$ is determined from the centroid potential of the mean force (PMF),

$$k_{\text{QTST}}(T;\xi^\ddagger) = \frac{4\pi R_\infty^2}{\sqrt{2\pi\beta\mu}} \exp\{-\beta[W(\xi^\ddagger)-W(0)]\}, \tag{7}$$

The term $W(\xi^\ddagger)-W(0)$ in the exponential is the free-energy difference between both dividing surfaces, usually obtained by using umbrella integration (UI) [32, 33] along $\xi$. And $\mu$ is the reduced mass of reactants. The transmission coefficient, $\kappa(t\to\infty;\xi^\ddagger)$, provides dynamical correction evaluated by the ratio between long-time limit and zero-time limit of the flux-side correlation function:

$$\kappa(t\to\infty;\xi^\ddagger) = \frac{c_{\text{fs}}^{(P)}(t\to\infty;\xi^?)}{c_{\text{fs}}^{(P)}(t\to 0_+;\xi^?)}, \tag{8}$$

which accounts for recrossing at the bottleneck ($\chi^\ddagger$) with ZPE.

The number of beads determines the accuracy of the simulation. More beads would lead more accuracy, but less computational efficiency. And there is also a formula to predict the minimum number of beads needed for a given system:

$$P_{\min} = \beta\hbar\omega_{\max}, \tag{9}$$

where $\omega_{\max}$ is the largest vibrational frequency of the system. In this work, it is the frequency of OH bond stretching, $3757$ cm$^{-1}$.[19] In each system at each temperature, $P_{\min}$ is calculated first and is rounded up to be power of 2. It worth to point out, at low temperatures, the number of beads needed would become large, making calculation time-consuming.

Another key parameter, the cross-over temperature, also needs considering with:

$$T_c = \hbar\omega_b/(2\pi k_B), \tag{10}$$

where $i\omega_b$ is the imaginary frequency of the reaction system at the bottleneck. The system below $T_c$ is considered at the so-called deep-tunneling region, where RPMD would give results with larger error, and much more number of beads than $P_{\min}$ needed to obtain results accurate enough.

The crossover temperature for the title reaction is $T_c = 326$ K with $\omega_b = 1422i$ cm$^{-1}$.[19]

The final rate coefficients should be corrected by a ratio of the electronic partition function

$$f(T) = \frac{Q_{\text{elec.}}^{\text{TS}}}{Q_{\text{elec.}}^{\text{reactants}}} = \frac{2}{2 + 2\exp(-\Delta E/T)}, \qquad (11)$$

where $\Delta E = 140$ cm$^{-1}$, to take account into the spin–orbit splitting of OH ($^2\Pi_{1/2}$ and $^2\Pi_{3/2}$)[34].

A proper umbrella force constant $k_l$ is highly needed to maintain the mean reaction coordinate $\xi_l$ properly oscillating near its reference value in the $l^{\text{th}}$ window, $\xi_l^0$. Especially, when the system in the region near TS, the gradient of potential energy becomes large, causing $k$ must be large enough to keep the configuration. But it cannot be set too large either, since there is also an upper bound of $k$ [33]:

$$\frac{k}{T} < \frac{9k_B}{(\Delta \xi)^2}, \qquad (12)$$

where $\Delta \xi$ is the width between neighboring windows. So, it should be $k < 0.28$ ($T$ / K) a.u. as $\Delta \xi = 0.01$, and $k < 1.14$ ($T$ / K) a.u. as $\Delta \xi = 0.005$.

In this work, to determine proper $k$ values, we set $k$ for each window as 0.1 ($T$/K) a.u. at the beginning and then increased it when one of the following criteria is satisfied: (1) The variance $\sigma > 5 \times 10^{-5}$, (2) The drift of mean value $|\xi - \xi^{\text{ref}}| > 10^{-3}$, (3) The ratio of the variance of the current trajectory to the variance of the last trajectory is greater than 1.1, the $k$ needs updating. Then we increase the force constant of the current window to its new value by multiplying a factor 1.2.

**2.2 Cayley Propagator**

The traditional propagator of RPMD is brought from the normal mode method in PIMD[35]. It was found not stable and ergodic in previous work[24, 36].

In the original path integral molecular dynamics (PIMD), the propagator making the system evolve from $t$ to $t+\Delta t$ is

$$\begin{bmatrix} q_i(t+\Delta t) \\ v_i(t+\Delta t) \end{bmatrix} = e^{L_i \Delta t} \begin{bmatrix} q_i(t) \\ v_i(t) \end{bmatrix}, \quad (13)$$

where $q_i(t) = \left(q_i^{(0)}(t), q_i^{(1)}(t), \cdots, q_i^{(P-1)}(t)\right)$ and $v_i(t) = \left(v_i^{(0)}(t), v_i^{(1)}(t), \cdots, v_i^{(P-1)}(t)\right)$ are the coordinates and momentum for $P$ beads respectively, and $L_i$ is the classical time evolution operator of the $i$ th atom. Using the Trotter expansion, the evolution induced by the free ring-polymer part, and the external potential part can be separated with an $\mathcal{O}(\Delta t^2)$ residual error,

$$e^{B\Delta t/2} e^{A\Delta t/2} e^{O\Delta t} e^{A\Delta t/2} e^{B\Delta t/2} \quad (14)$$

where $A$, $B$ and $O$ are the time evolution operator of pure free ring-polymer, the time evolution operator of the system on potential energy surface, and thermostat. This expansion scheme is denoted as "BAOAB", and is recommended in a recent study.[37]

A recent work[24] revealed that such a "BAOAB" integrator is not strongly stable with respect to the external potential, which causes poor ergodicity when using a large timestep and a large number of beads.

The "BCOCB" integrator is then introduced in the following study[25] to solve this issue by replacing the operator "A" from the exponential propagator to the Cayley propagator:

$$\exp(\Delta t A / 2) \rightarrow [\text{cay}(\Delta t A)]^{1/2}, \quad (15)$$

where

$$\text{cay}(\Delta t A) = \left(I - \frac{1}{2}\Delta t A\right)^{-1} \left(I + \frac{1}{2}\Delta t A\right).$$

Such "BCOCB" integrator is shown to be not only strongly stable with respect to the external potential but also provides the exact equilibrium position marginal distribution in the phase space for harmonic external potentials. It is later shown that the "BCOCB" integrator is the unique one in the "BAOAB" family that can achieve such good properties.[36]

As mentioned above in equation (3), (7) and (8), one propagation scheme in two different types is merged into RPMD calculation. In the first period of calculating $k_{\mathrm{QTST}}(T;\xi^{\ddagger})$, thermostat PIMD, the BCOCB is used to calculate PMF. In the second period of calculating the transmission coefficient, RPMD, the BCB is used to calculate the recrossing coefficients. The Cayley propagator has been utilized in a recent work about RPMD in gas-surface reactions.[38]

2.3 Minimum Reaction Path

In the period of calculating $k_{\mathrm{QTST}}(T;\xi^{\ddagger})$, the reference reaction coordinates $\chi_i^{\mathrm{ref}}$ in each UI window $i$ should be specified before evaluating the PMF according to eq.(7). In the RPMD rate coefficient calculations, it is determined by starting a short classical MD from the TS to both of the asymptotic regions, and then monitoring the reaction coordinate to choose the proper initial configuration within in each window. This protocol can hardly make the RCs located near the minimum of the biased PES $(V(\boldsymbol{q})+\frac{1}{2}k_{\mathrm{f}}(\xi_i(\boldsymbol{q})-\xi_i^{\mathrm{ref}})^2)$ for each window, leading to a long time of equilibrium when doing umbrella sampling. This shortage would consume much computational time in calculation of multi-atomic reactions. For dealing with that, in this work a method named optimization-interpolation method (OIM), successfully applied to a multi-channel reaction recently,[39] was exploited to locate the exact correspondence to the RCs and Cartesian coordinates. The OIM calculation is divided into two stages. The Cartesian coordinates of the reaction complex in each window were first optimized on the biased PES from the TS to each side. The RCs would drift slightly to the negative direction of gradient along the RCs. After generating all Cartesian coordinates in each window, a cubic spline interpolation was performed on each component of Cartesian coordinates. Hence, the Cartesian coordinates were constructed as a smooth function of RCs. After OIM calculation, the minimum reaction path (MRP), linking the reactant and product through the TS, is generated.

**2.4 Computational Details**

The globally full-dimensional PIP-NN PES with high accuracy developed by Li *et al.*[19] (PIP-NN PES) was used in our RPMD calculations. It's constructed based on about 135,000 points at the level of UCCSD(T)-F12a/AVTZ, whose root-mean-square fitting error is less than 0.09 kcal/mol.

All simulations about rate coefficients reported here used RPMDrate[30] and with our customizations.

The number of beads used in simulations is based on Eq.(9), and rounded up to the power of 2. During the sampling, the force constants needed for windows near the transition state ($0.9 \leq \xi \leq 1.05$) would exceed the criteria shown in Eq.(12) with $\Delta\xi = 0.01$. So, we reduced the size of windows in that region ($\Delta\xi = 0.005$). In each window, the system was equilibrated for 0.2 ps, following a production run of 10 ps per trajectory. Here 30 trajectories were employed in each window and equivalent to 300 ps for the system totally. Such choice is a balance between accuracy and efficiency, and the convergence validation is shown in SI. The Andersen thermostat[40] was used in all NVT ensemble simulations. The time step used for traditional ring-polymer propagator at $T > 500K$ is 0.1 fs, 200 K to 300 K is 0.2 fs, and for Cayley propagator is 0.5 fs.

In the calculation of transmission coefficients, a long parent trajectory constrained at the peak of PMF was simulated 10 ns using the SHAKE algorithm, with the purpose of generating initial configurations that were sampled every 0.2 ps. For each of these constrained configurations, 100 child trajectories were then spawned at the random initial momentum from the Boltzmann distribution. Each child trajectory was then propagated without constraint for 0.5 ps to reach the "plateau" time.

At each temperature, the RPMD calculations were firstly performed with one bead, which provides the classical limit, and then with a sufficient number of beads ensuring convergence.

The minimum energy paths (MEPs) for the title reaction with each isotope are prepared by using POLYRATE[41] on the PIP-NN PES, searched downhill from the transition-state geometry to both reactant and product sides in mass-weighted reaction coordinates with a step size of $5 \times 10^{-4}$ amu$^{1/2}$Bohr. Then the MEPs were analyzed from $s = -10$ to 10 amu$^{1/2}$Bohr. The Hessian matrix was updated per 10 points. All calculations performed in mass-scaled coordinates with different reduced masses. The vibrational adiabatic potentials $V_a^G(s) = V_{MEP}(s) + V_{ZPE}(s)$ along the MEP were also obtained using POLYRATE. It is the vibrational zero-point energy of the generalized normal-mode analysis. [20]

We have also compared the reaction path with that on PES-2000 and PES-2014, as shown in Figure S7. Both PES-2000[42, 43] and PES-2014[23] have shallower vdW well and higher potential barrier than PIP-NN PES[19]. In general, the vdW well results in complex forming near TS, so that leads smaller rate coefficients. The higher barrier also leads smaller rate coefficient. From the results of rate coefficients, one can observe the too shallow vdW is the major deficiency of both PES-2000 and PES-2014, since the rate coefficients calculated on them are artificially larger than experimental values at temperatures lower than 300 K.

### 4. Results and Discussion

The minimum energy path (MEP) and the ground state adiabatic energy $V_a^G$ for all four isotopic reactions are shown in Figure 1. Here only the regions near the TS are shown for clarity, and the full-range MEPs can be found in Figure S5. For MEP, all the 4 reactions are exothermic and with highly similar shape, since in this plot, the mass-scaled reaction coordinate[44] is used, and the reduced mass of all reactions ( $\mu_{OH+CH_4}=8.25$, $\mu_{OD+CH_4}=8.48$, $\mu_{OH+CD_4}=9.20$, and $\mu_{OH+^{13}CH_4}=8.51$) are nearly the same. Only at higher temperature can the TST give reasonable results.[45] But from the view of skewing angles, considering it for all the title reactions are very small, say, 19.9° for OH + CH$_4$, 19.6° for OD + CH$_4$, 26.3° for OH + CD$_4$, and 19.5° for OH + $^{13}$CH$_4$, suggesting all of them would be with large quantum effects, such as tunneling, kinetic crossing and resonance, causing the TST results smaller than experimental data at lower temperatures ($T < 300K$)[45]. On the other hand, for $V_a^G$, the heavier reactant gives the lower value, due to the vibrational frequency contributing to the zero point energy (ZPE) reduction from OH (3757 cm$^{-1}$) to OD (3198 cm$^{-1}$), and CH$_3$ (3067 cm$^{-1}$) to CD$_3$ (2188 cm$^{-1}$). The MEPs for three isotopes (OH + CH$_4$, OH + CD$_4$, and OD + CH$_4$) can also be found in previous work by J. Li and H. Guo.[19]

**4.1 OH + CH$_4$**

The rate coefficients for the OH + CH$_4$ reaction and its isotopes, calculated by RPMD and other theoretical methods, and measured from experiments, are summarized in Table 1. The rate coefficients for the reaction OH+CH$_4$ are also depicted in Figure 2, with results from other theoretical methods and experimental results. In detail, all collected experimental results are plotted in Figure 2(a), and the theoretical ones are shown in Figure 2(b). First of all, at low temperature, the reactants are trapped in the vdW well at first, forming a complex with orientation favoring to the reaction, leaving the rate coefficients larger than those predicted by Arrhenius' law.

Another feature of the title reaction is that the recrossing is significant in all the four isotopic reactions, which can be observed in the table. This phenomenon would be attributed to the small skew angle of the title reaction. Here the word recrossing is used loosely since it contains not only the classical recrossing over the barrier, but also the quantum tunneling effect which is reflected by the smaller transmission coefficient value at lower temperature. This is also the characteristic feature of the heavy-light-heavy reactions, as discovered in O/Cl+CH$_4$ reactions[46, 47]. Furthermore, the small skewing angle would also lead to a quantum reactive scattering resonance[47], which would be another source of deviation between RPMD results and experimental one. Unfortunately this system is too large for time-dependent quantum dynamics calculations, it cannot elucidate the effect of quantum resonance in the title reactions till now. Another interesting feature of the RPMD results is the position of the peak of PMF ($x^{\ddagger}$), which is increasing with temperature, in contrast to the reactions of Cl/O/H+CH$_4$[46-48]. This can be understood as a significant van der Waals (vdW) well on PES before the TS in the reactant region. This trend is also evidenced in the reactions of HCl + OH[49] and S + H$_2$[50].

From both the Table 1 and Figure 2, the RPMD rate results in this work agree perfectly with experimental results among the entire range of temperatures investigated, 200 K to 1500 K. Considering when the temperature is below $T_c$ (326 K for title reaction), the RPMD would overestimate the rate coefficients due to the asymmetric barrier[51], our results coincide with this prediction. Even the error becomes larger at 200 K, the relative error to the experimental values are still less than 20%, proving again RPMD is a highly reliable method for obtaining rate coefficients.

The RPMD results based on other PESs have great errors when the temperature goes below $T_c$. Such as, on the PES-2000[3] (RPMD-2000) at 200 K is an order of magnitude larger than the experimental value. Results from both RPMD (RPMD-2014) and quantum instanton (QI) on PES-2014[23] are slightly better than that those obtained on PES-2000, but there is still an error of 5 times at 200 K. At 300 K, near $T_c$, the result from RPMD-2000 is still an order of magnitude larger than the experimental value. At this temperature, RPMD-2014 and QI results matched the experimental value, but that would due to the error cancellation. When the temperature is greater than 500 K, the RPMD-2000 and QI results are getting closer to the experimental values, but that from RPMD-2014 is lower than the experimental values.

To better understand the above results, it's helpful to recall the details of other PESs used in previous works. The PES-2014 is fitted thoroughly from *ab initio* points to an existed functional form, but the PES-2000 is constructed with both *ab initio* points and experimental data.[43] On the contrary, the PIP-NN PES is obtained from *ab initio* points at CCSD(T)-F12a/AVTZ level and is free from the deficiency of functional form with the help of neural network. Although the heights of the energy barrier from the asymptotic region to TS are nearly the same among PES-2000 (6.6 kcal/mol), PES-2014 (6.4 kcal/mol) and PIP-NN PES (6.3 kcal/mol), the heights from the vdW well to TS differ larger, 7.0 on PES-2000 and 6.8 kcal/mol on PES-2014 versus 7.53 on PIP-NN PES[19]. Such difference would be the source making rate coefficients from both PES-2000 and PES-2014 are artificially higher than the experimental value at lower temperatures. This difference can also explain why at higher temperature, the results from all three PESs are nearly the same, since the higher temperature corresponds to higher kinetic energy, which smears out the vdW well and the absolute height of barrier plays the determinate role.

The topology of PES is also important since it governs the frequencies of reaction species, which can be reflected by $T_c$, since it is proportional to the imaginary frequency of the TS. On different PESs, $T_c$ is 390 K for PES-2000, 370 K for PES-2014, and 326 K for PIP-NN PES. Such deviation would contribute to the large deviation of results on PES-2000 and PES-2014 from the experimental values.

Figure 3 shows the potential of mean force (PMF) of RPMD rates at temperatures 200 to 1500 K, and the subplots highlight the position of peaks. All the PMF curves from reactants to TS show similar tendencies. With the rising temperature, the barrier heights become higher with the increasing kinetic energy, and the vdW well on the left of TS becomes filled out by the kinetic energy correspondingly. The RPMD barrier heights for title reaction at 200, 300, 500, 1000 K are 5.58, 6.90, 9.11, 13.63 kcal/mol with the $\xi^{\ddagger}$ as 0.971, 0.983, 0.993, 1.001, respectively. The reaction coordinates of barriers $\xi^{\ddagger}$ drift right when the temperature goes up, as shown in the Table 1, mentioned above.

The transmission coefficients as a function of time are shown in the right panel of Figure 3. From the plots, the "plateau" values of $\kappa(t)$ are relatively small as mentioned above. Especially at low temperatures, the plateau values are 0.383 for 200 K and 0.482 for 300 K, similar as observed in a previous study for H + CH$_4$.[48] All the transmission coefficients converged to their plateau values after a short period of fluctuation, which is from the choice of the dividing surfaces, and with less physical meaning[47].

The single-bead (classical) and multi-bead (RPMD) PMF for title reaction at temperatures 200 K, 300 K, 500 K, and 1000 K are shown in Figure 4, left panel. Unlike obtained from RPMD, the position of peaks of classical PMFs are almost unchanged with the temperature, $\xi^{\ddagger(cl)} = 1.001$ for 200 and 300 K, and $\xi^{\ddagger(cl)} = 1.003$ for 500 and 1000 K. The difference between $x^{\ddagger}$ from classical and RPMD reduced from 0.030 at 200 K to 0.002 at 1000 K with the temperature rising, combined with the difference between heights of both peaks $\Delta\Delta G(x^{\ddagger}) = \left| \Delta G^{(\text{RPMD})}(x^{\ddagger}) - \Delta G^{(\text{cl})}(x^{\ddagger}) \right|$ is reducing with increasing of temperature, means the results are closer to the classical limit. Figure 4 also shows comparison of the $\kappa$ from RPMD and from classical MD in the right panel. $\kappa$ from RPMD is smaller than its one-bead counterparts. The difference between $\kappa$ from single-bead and converged number of beads $\Delta\kappa = \left|\kappa^{\text{RPMD}} - \kappa^{\text{cl}}\right|$ becomes smaller with the temperature rising, indicating again the reaction reaches its classical limit. The transmission coefficients display opposite trends from classical MD and RPMD, which decreases with temperature from classical MD, but increases with temperature from RPMD. That means for the title reaction, from classical MD the recrossing events increase with temperature caused by thermal motion, while from quantum mechanics, the recrossing becomes frequent when the temperature goes down, caused by tunneling.

In this work, the rate coefficients of the title reaction were also calculated using the Cayley propagator at 200, 300, 500 K. The time step is set to 0.5 fs, which is five times larger than the traditional time step of the original propagator. That means the CPU time is only one fifth using this propagator. The results from Cayley propagator match both the experimental and RPMD rate values well, with the maximum relative error no more than 10%. So Cayley propagator RPMD shows its power of obtaining rate coefficients with both high accuracy and high efficiency.

The comparison of PMF and transmission coefficients of original and Cayley propagator for the title reaction, OH + $CH_4$, has been shown in figure S7. The results of these two propagators are in great agreement at all three temperatures. At 200 K, the PMF from Cayley propagator RPMD is lower than that from standard RPMD, but $\kappa$ is also lower. At last, the deviations canceled out, leaving final result from Cayley propagator RPMD at 200 K consistent with that from standard RPMD.

We have validated the convergence of RPMD rate calculations,[52] and found 30 trajectories per each umbrella window are enough to give converged PMF, as shown in Figure S4. Also, we have validated the convergence of results with the number of beads. At 200 K and 500 K, we have used two numbers of beads, 32 and 48 for 200 K and 16 and 32 for 500 K, to do the RPMD calculations using Cayley propagator. The minimum numbers of beads from formula (9) are 27 and 11 separately. The deviations of rate coefficients from different bead numbers are only 13% at 200 K and 5% at 500 K. The detailed plot of the comparison between PMF and transmission coefficient curves are shown in Figure S8.

To clarify the efficiency of Cayley propagator, we have also performed two simulations with 0.5 fs time interval using original propagator, at 1500 K and 200 K separately. The obtained rate coefficients are largely lower than that in 0.1 fs, with relative errors of rate coefficients as −67% at 1500 K and −43% at 200 K, respectively.

**4.2 OD + CH$_4$**

The rate coefficients of RPMD and other theoretical and experimental values are drawn in Figure 5. The RPMD rate coefficient at 300 K matches well with the experimental value, with an error of about 17%[13, 53]. But the rates on the whole range of temperature calculated on PES-2014 using RPMD have large deviation, which is higher than the experiment at lower temperatures but lower than all other theoretical values. The rate coefficients obtained from canonical variational transition-state theory on microcanonical optimized multidimensional tunneling (CVT/μOMT) on PIP-NN PES agree well with the experimental value at about 250 K to 350 K only. Results from CVT/μOMT are overestimated at higher temperatures while underestimated at lower temperatures. On the other hand, results from RPMD are in great agreement with the experiments in the whole temperature range, just like in the case of OH+CH$_4$.

Figure 6 collects the KIE of OH/OD+CH$_4$ in the upper panel. Results from RPMD on PIP-NN PES agree excellently with the available experimental values, even at low temperatures. The RPMD KIEs of OH/OD+CH$_4$ in this work is 0.83, 0.92, 0.96, and 1.03 at 200 K, 300 K, 500 K, 1000 K, respectively. The RPMD results on PES-2014 couldn't describe these quantum effects well, which led to great error at the whole temperature range. The results from CVT/μOMT on PIP-NN PES display oscillation at the temperature range from 500 K to 1000 K, and show deficiency at temperatures lower than 500 K. These results reflect that VTST cannot describe the title reaction correctly. RPMD results in this work show abnormal KIE at temperature less than 700 K, and normal KIE at higher temperatures. Since there is no experimental result at temperatures larger than 700 K, we hope future experimental works confirm our finding.

**4.3 OH + CD$_4$**

The isotope reaction OH + CD$_4$ was also investigated to provide a thorough understanding of the mechanism. The rate coefficients and KIEs were collected in the lower panel of Figure 5 and Figure 6, respectively. The Arrhenius curve of reaction OH + CD$_4$ matches the experimental value excellently, even better than that of OH + CH$_4$ and OD + CH$_4$. The results from other theoretical methods can also provide reasonable values, too. This agreement would be attributed to that CD$_4$ has smaller quantum effects due to the heavier mass of D than H. The RPMD KIEs in this work is 17.48, 8.19, 4.41, and 2.07 at 200 K, 300 K, 500 K and 1000 K, respectively. KIE from RPMD in this work agree best to experimental values among theoretical results shown, especially at 300 K. However, the RPMD rates based on the PES-2014 overestimate much the results at temperature lower than 400 K. The rate coefficients and KIEs from CVT/μOMT on PIP-NN PES also near the experimental results in general, but systematically lower than experimental values, might stem from lack of recrossing in VTST methods or other error accumulation.

## 4.4 OH + $^{13}CH_4$

The rate coefficients of isotope of carbon, OH + $^{13}CH_4$, are obtained from the Cayley propagator RPMD and CVT/μOMT on PIP-NN PES. At the experimental temperature, 273 K, 293 K, and 353 K, the KIEs from Cayley propagator RPMD are 1.0178, 0.9970, and 1.0070, whose average 1.0073 is in good agreement with the experimental value of 1.0054 ± 0.0009 over the temperature range of 273−353 K. KIEs from CVT/μOMT in this work are 1.0306, 1.0100, and 0.9964 respectively. The average value of 1.0123 of them is close but larger than the experimental value.

The $^{12}C/^{13}C$ KIEs for OH + $CH_4$ is a stringent test, whose results depend strongly on the treatment of the torsional anharmonicity of the lowest vibrational mode.[23] The Cayley propagator RPMD includes all vibrational degree of freedom, which correctly accounts for all the anhamonicity effects. Such great agreement shows again that the accurate rate coefficients must be obtained using both RPMD and high-fidelity PES.

## 5. Conclusion

We investigated the rate coefficients of the title reaction, OH + $CH_4$, and three of its isotope reactions, OD+ $CH_4$ , OH + $CD_4$, and OH + $^{13}CH_4$, using the RPMD with Cayley propagator, after validation in the reaction of OH + $CH_4$. The RPMD is expected to the best semiclassical way to calculate rate coefficients. Especially, for reactions with large number of DOF, it's the only accurate way till now. Based on a high-fidelity PIP-NN PES, the RPMD rate results are in great agreement with the experiment values. Even in deep-tunneling region below the crossover temperature 326 K, the deviation of rate values from RPMD is still less than 20% relative to the experiment values, showing great power of RPMD.

The kinetic isotope effects of deuterated reactants OD and $CD_4$ are also investigated. It is valuable to explore the quantum effects using KIE, but the traditional TST and its extensions will lead to a huge error. Our KIE using RPMD are also in good agreement with the experimental values.

The utilizing of ultra-stable and symplectic propagator, Cayley propagator, shows considerable speed up of the calculations without accuracy loss, with deviation less than 10% comparing with the traditional propagator RPMD. With such efficiency, we obtained successfully the KIE of $^{12}C/^{13}C$.

Nowadays, more and more high-fidelity PESs have been developed.[54-57] We hope that the Cayley propagator RPMD can be further used in more complicated reactions with even larger degrees of freedom.

**Author Contributions**

Xiongfei Gui and Wenbin Fan contributed equally to this work.

**Notes**

The RPMD code with Cayley propagator and other unpublished codes are available from the corresponding author with reasonable requests.

**Supporting Information**

The parameters for RPMDrate calculations.

The minimum reaction path in RPMD reaction coordinates.

The contour plot of the title reaction along two reactive bonds.

The umbrella force constants for all isotopic reactions.

The convergence with the number of trajectories.

The minimum energy path for all isotopic reactions.

The PMFs and transmission coefficients of standard and Cayley propagator.

The potential energy profile along reaction coordinate of PIP-NN PES and PES-2014.

The validation of convergence with the number of beads.

**Acknowledgment**

This work was supported by the National Nature Science Foundation of China (No. 22173057), the research grant (No. 21JC1402700) from Science and Technology Commission of Shanghai Municipality, and the Key Research Project of Zhejiang Laboratory (No. 2021PE0AC02). We appreciate the High Performance Computing Center "Ziqiang 4000" of Shanghai University. We also gratefully acknowledge HZWTECH for providing computation facilities. The authors thank Jun Li in Chongqing University for providing a high-fidelity PIP-NN PES program. The authors also thank helpful discussions from Thomas F. Miller III in California Institute of Technology and Hua Guo in University of New Mexico.

**Table 1**. Results from RPMD calculations of the rate coefficients for the title reaction and its isotopic reaction. The rate coefficients are in cm$^3$ molecules$^{-1}$ s$^{-1}$, $\Delta G$ is in kcal/mol, $\xi^{\ddagger}$ and $\kappa$ are dimensionless.

| $T$/K | 200 | 300 | 500 | 700 | 1000 | 1500 |
|---|---|---|---|---|---|---|
| $P$ [a] | 32 | 32 | 16 | 16 | 8 | 4 |
| $f(T)$ [b] | 0.732 | 0.662 | 0.599 | 0.571 | 0.550 | 0.533 |
| OH + CH$_4$ | | | | | | |
| $\xi^{\ddagger}$ | 0.971 | 0.983 | 0.993 | 0.997 | 1.001 | 1.004 |
| $\Delta G$ | 5.58 | 6.90 | 9.11 | 10.95 | 13.63 | 17.46 |
| $k_{QTST}$ | 1.79×10$^{-15}$ | 2.61×10$^{-14}$ | 3.70×10$^{-13}$ | 1.60×10$^{-12}$ | 5.29×10$^{-12}$ | 1.76×10$^{-11}$ |
| $\kappa$ | 0.383 | 0.482 | 0.539 | 0.555 | 0.582 | 0.581 |
| $k_{RPMD}$ | 5.05×10$^{-16}$ | 8.37×10$^{-15}$ | 1.20×10$^{-13}$ | 5.08×10$^{-13}$ | 1.70×10$^{-12}$ | 5.46×10$^{-12}$ |
| $k^{cl}$ [c] | 2.21×10$^{-19}$ | 8.36×10$^{-17}$ | 1.29×10$^{-14}$ | 1.32×10$^{-13}$ | 8.82×10$^{-13}$ | / |
| $k_{CVT/\mu OMT}$[20] | 2.34×10$^{-16}$ | 5.80×10$^{-15}$ | 1.21×10$^{-13}$ | 5.43×10$^{-13}$ | 2.12×10$^{-12}$ | 9.82×10$^{-12}$ |
| $k_{RPMD}$ on PES-2014[23] | 1.54×10$^{-15}$ | 8.15×10$^{-15}$ | 6.23×10$^{-14}$ | 2.31×10$^{-13}$ | 9.63×10$^{-13}$ | / |
| $k^{exp}$ | 4.00×10$^{-16}$ [13] | 7.66×10$^{-15}$ [58] | 1.01×10$^{-13}$ (498 K)[11] | 4.78×10$^{-13}$ (698 K)[6] | 2.16×10$^{-12}$ (1005 K)[6] | 6.24×10$^{-12}$ [17] |
| OH + CH$_4$ with Cayley propagator | | | | | | |
| $\xi^{\ddagger}$ | 0.965 | 0.983 | 0.993 | | | |
| $\Delta G$ | 5.46 | 6.93 | 9.21 | | | |
| $k_{QTST}$ | 2.43×10$^{-15}$ | 2.46×10$^{-14}$ | 3.35×10$^{-13}$ | / | / | / |
| $\kappa$ | 0.299 | 0.483 | 0.535 | | | |
| $k_{RPMD}$ | 5.33×10$^{-16}$ | 7.85×10$^{-15}$ | 1.08×10$^{-13}$ | | | |
| rel. error | 6% | −6% | −10% | | | |
| OD + CH$_4$ | | | | | | |
| $\xi^{\ddagger}$ | 0.971 | 0.982 | 0.992 | 0.996 | 1.000 | 1.003 |
| $\Delta G$ | 5.49 | 6.82 | 9.02 | 10.97 | 13.49 | 16.79 |
| $k_{QTST}$ | 2.23×10$^{-15}$ | 2.90×10$^{-14}$ | 3.99×10$^{-13}$ | 1.56×10$^{-12}$ | 5.60×10$^{-12}$ | 2.17×10$^{-11}$ |
| $\kappa$ | 0.371 | 0.474 | 0.522 | 0.553 | 0.566 | 0.598 |
| $k_{RPMD}$ | 6.06×10$^{-16}$ | 9.09×10$^{-15}$ | 1.25×10$^{-13}$ | 4.94×10$^{-13}$ | 1.75×10$^{-12}$ | 6.94×10$^{-12}$ |
| KIE | 0.83 | 0.92 | 0.96 | 1.03 | 0.97 | 0.79 |
| KIE on PES-2014[23] | / | 0.67 | 0.76 | 0.89 | 1.02 | / |
| OH + CD$_4$ | | | | | | |
| $\xi^{\ddagger}$ | 0.984 | 0.992 | 0.997 | 1.000 | 1.001 | 1.005 |
| $\Delta G$ | 6.86 | 8.16 | 10.65 | 12.56 | 15.13 | 18.09 |
| $k_{QTST}$ | 6.84×10$^{-17}$ | 2.97×10$^{-15}$ | 7.42×10$^{-14}$ | 4.79×10$^{-13}$ | 2.35×10$^{-12}$ | 1.35×10$^{-11}$ |
| $\kappa$ | 0.553 | 0.604 | 0.636 | 0.652 | 0.653 | 0.659 |
| $k_{RPMD}$ | 2.78×10$^{−17}$ | 1.19×10$^{−15}$ | 2.84×10$^{−14}$ | 1.79×10$^{−13}$ | 8.47×10$^{−13}$ | 4.75×10$^{−12}$ |
| KIE | 17.48 | 8.19 | 4.41 | 2.76 | 2.07 | 1.46 |
| KIE on PES-2014[23] | 39.71 | 13.12 | 3.36 | 2.16 | 1.61 | / |
| KIE from QI[d,3] | 96.39 | 16.81 | 4.45 | 3.12 | 2.25 | / |

a: Number of beads.

b: $f(T)$ is defined in the equation (11).

c: Single-bead RPMD rates. "cl" means classical.

d: QI means quantum instanton.

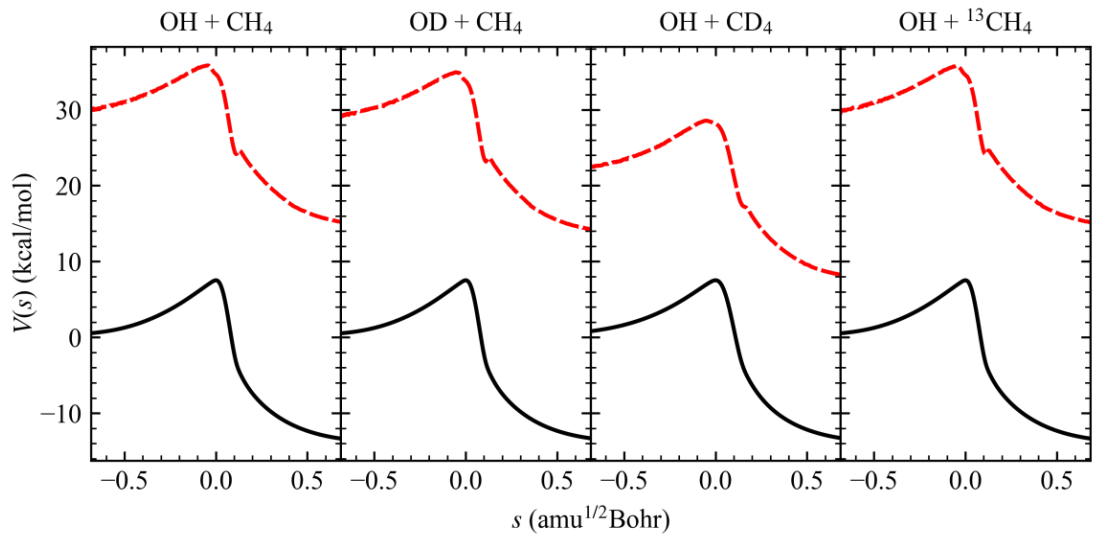

**Figure 1.** The minimum energy path (MEP, black solid line) and the ground vibrational state adiabatic potential ($V_a^G$, red dashed line) for all the four systems on PIP-NN PES along the mass-weighted reaction coordinate used in POLYRATE[41]. –

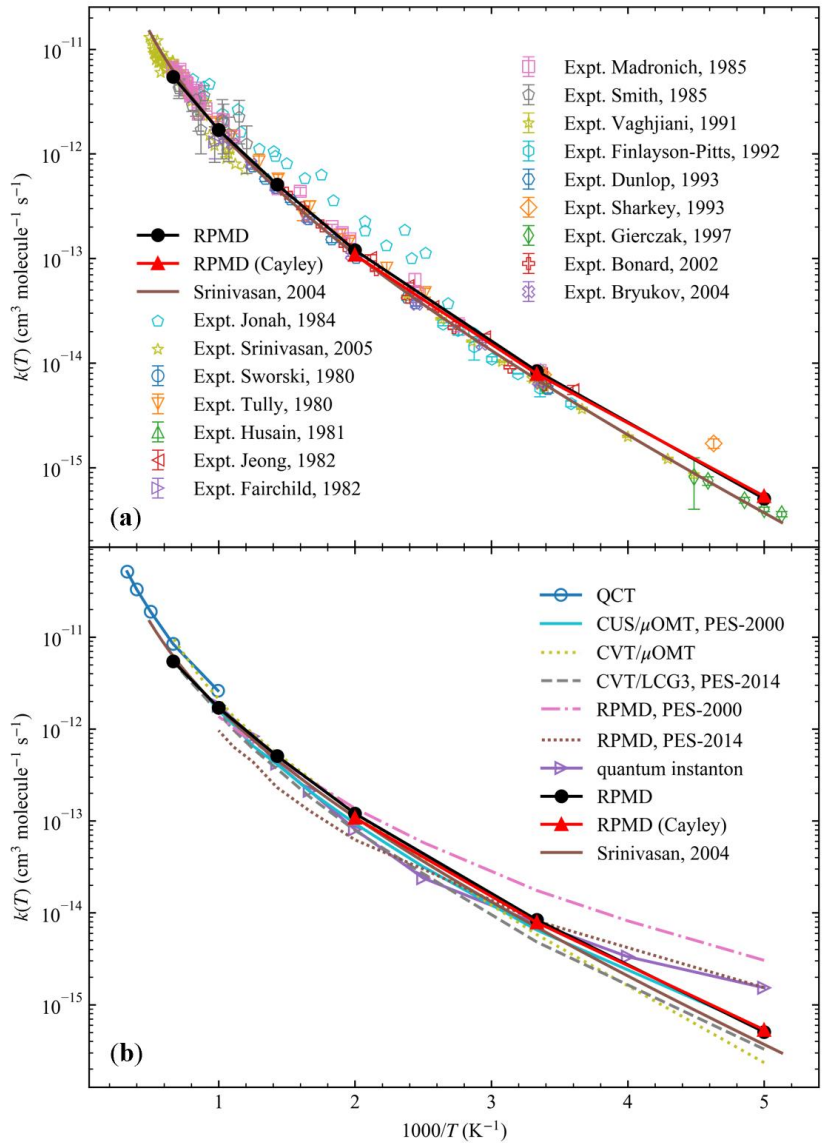

**Figure 2.** Comparison of rate coefficients of OH+CH$_4$ from traditional RPMD and Cayley RPMD on the PIP-NN PES. For comparison, other (a) experimental[4-7, 9-13, 15-17, 58-61] and (b) theoretical[20, 23, 42, 62] results are also shown. Especially, in subplot (b) the experimental results collected by Srinivasan and fitted to an empirical formula $k=1.66\times10^{-18}T^{2.182}\exp(-1231\,\text{K}/T)$ [17] is also included.

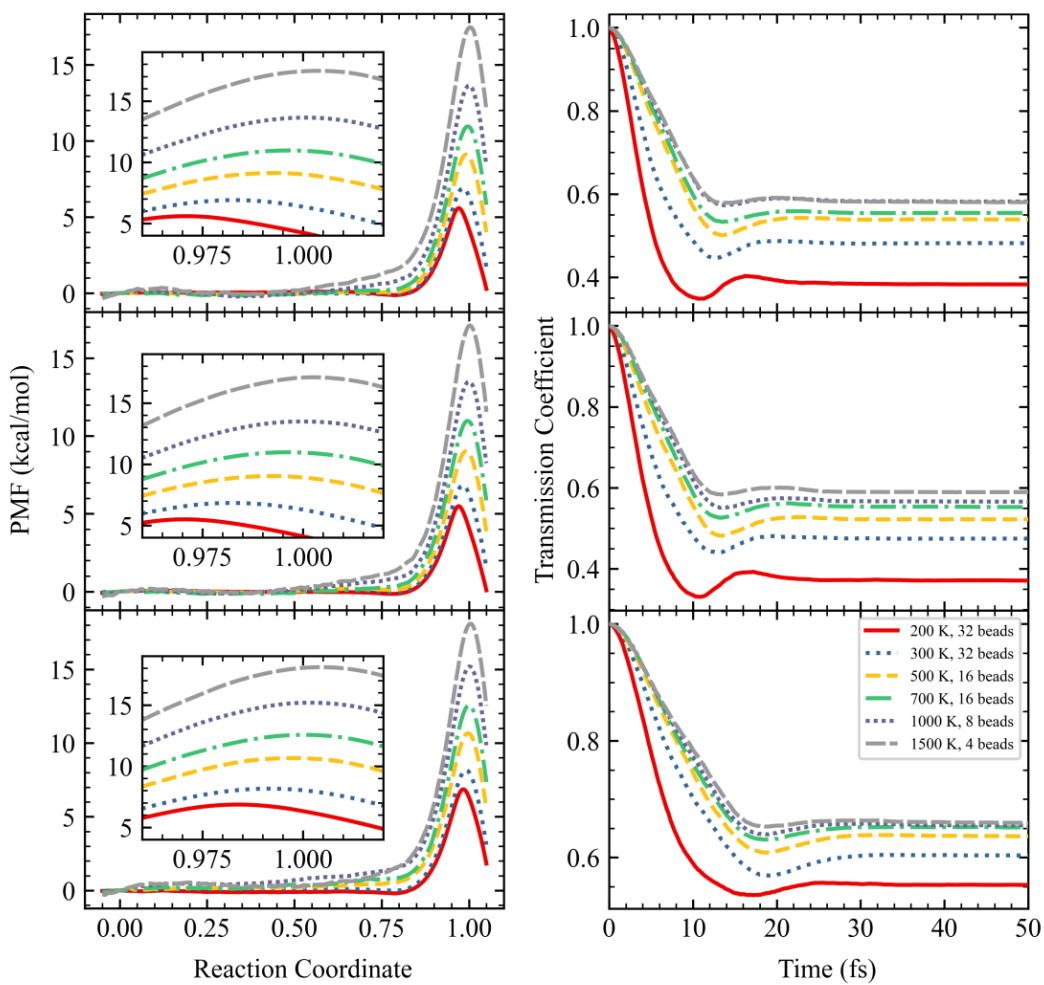

**Figure 3.** The potential of mean force (PMFs, left panel) and transmission coefficient (right panel) for OH + $CH_4$, OD + $CH_4$, OH + $CD_4$ (from top to bottom) at 200 K, 300 K, 500 K, 700 K, 1000 K, and 1500 K on the PIP-NN PES. The dots in each insect PMF plots represent the peak of PMF, and the vertical and horizontal dashed lines represent the corresponding reaction coordinate and PMF values.

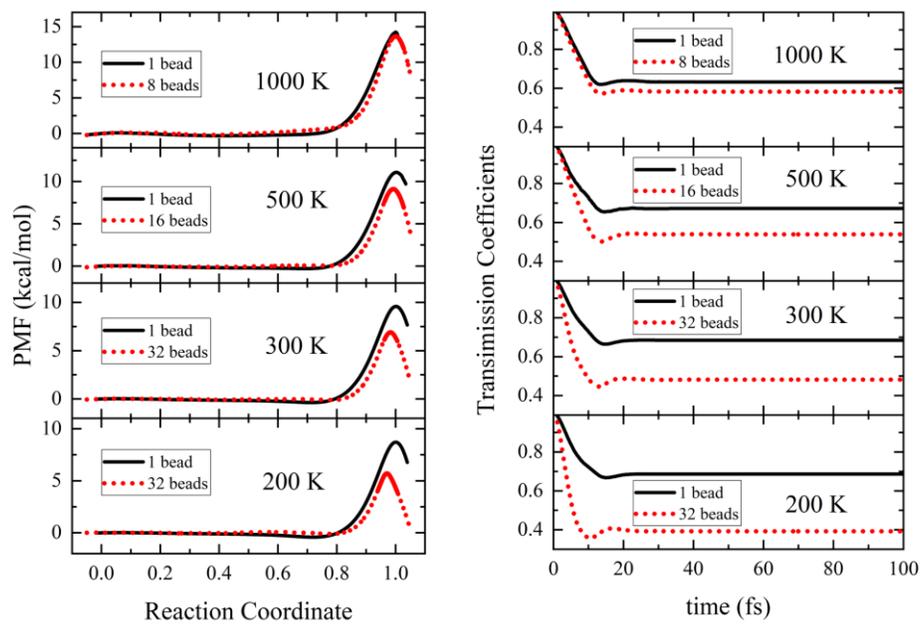

**Figure 4.** A comparison with the PMF profiles from converged RPMD and from classical (single-bead) MD (left panel) and corresponding time dependence of transmission coefficients (right panel) at 200 K, 300 K, and 500 K of the reaction OH + $CH_4$ on the PIP-NN PES.

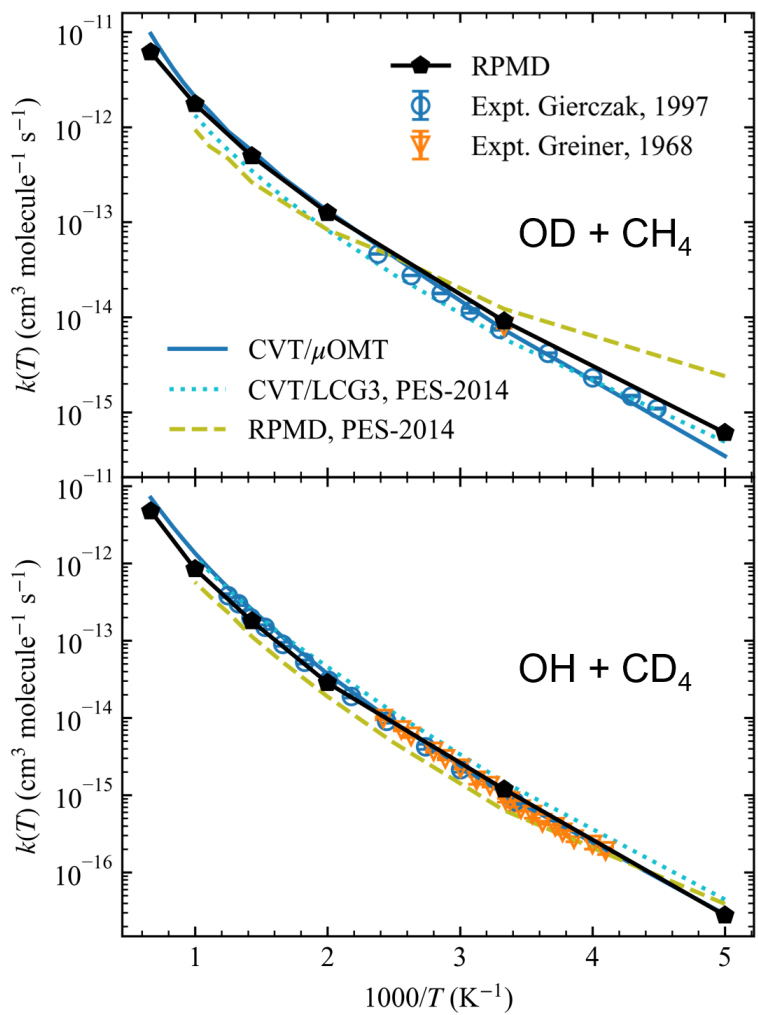

**Figure 5.** Comparison of rate coefficients of OD + CH$_4$ (upper panel) and OH + CD$_4$ (lower panel), from Cayley RPMD and CVT/μOMT[20] on the PIP-NN PES, and CVT/LCG3 and traditional RPMD on PES-2014[23]. For comparison, experimental counterparts[13, 53] are also included in both panels.

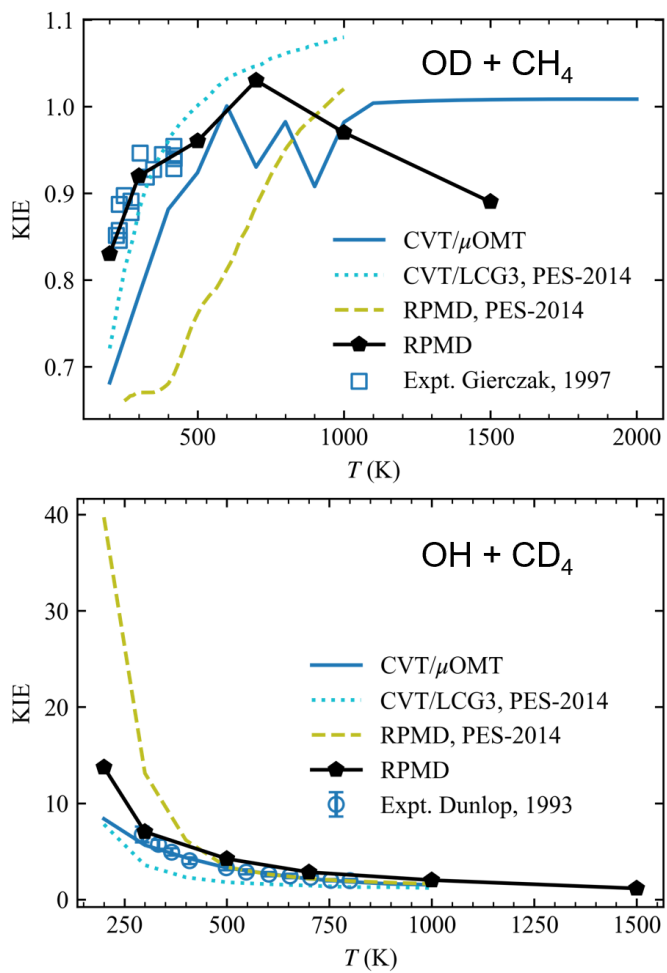

**Figure 6.** Comparison between the calculated kinetic isotope effect (KIEs) of OD + CH$_4$ (upper panel) and OH + CD$_4$ (lower panel) in this work, previously calculated[20, 23] and experimental ones[11, 13].

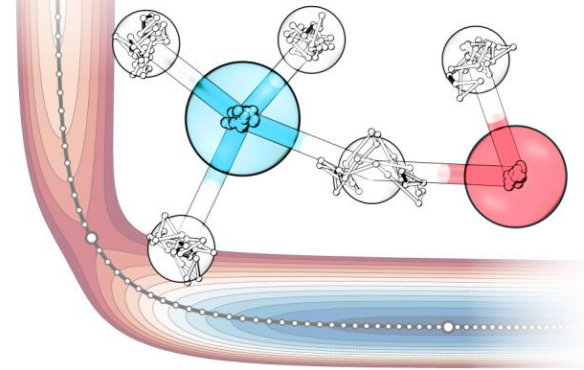

TOC